\documentclass[12pt,a4paper]{cibb}

\usepackage{subfigure}
\usepackage{graphicx}
\usepackage{amsmath,amsfonts,latexsym,amssymb,euscript,xr}
\usepackage{booktabs}
\usepackage[nodayofweek]{datetime}
\usepackage{hyperref}

\usepackage[table]{xcolor}
\usepackage{color,colortbl,tabularx}

\usepackage[english]{babel}
\usepackage[protrusion=true,expansion=true]{microtype}
\usepackage{amsmath,amsfonts,amsthm}
\usepackage{pifont}

\definecolor{LightBlue}{rgb}{0.88,0.9,0.9}

\title{\Large $\ $\\ \bf Comparing Propensity Score-Based Methods in Estimating the Treatment Effects: A Simulation Study}

\author{\large Sara Poletto$^1$, Enrico Longato$^{2}$, Erica Tavazzi$^{3}$, and Martina Vettoretti$^{*,4}$}
\address{\footnotesize $\ $\\$^1$ Department of Information Engineering, Università degli Studi di Padova, Padua, Italy. sara.poletto.1@phd.unipd.it, 0009-0004-4967-3051 \\
$^2$ Department of Information Engineering, Università degli Studi di Padova, Padua, Italy. enrico.longato@unipd.it, 0000-0001-5940-645X \\
$^3$ Department of Information Engineering, Università degli Studi di Padova, Padua, Italy. \\ erica.tavazzi@unipd.it, 0000-0001-6188-6413 \\
$^4$ Department of Information Engineering, Università degli Studi di Padova, Padua, Italy. martina.vettoretti@unipd.it, 0000-0002-5020-1818 \\
\bigskip
$^*$corresponding author
}

\abstract{\small propensity score, confounding, treatment effect, observational study, simulation 
\normalsize
\\[17pt] 
{\bf Abstract.} In observational studies, the recorded treatment assignment is not purely random, but it is influenced by external factors such as patient characteristics, reimbursement policies, and existing guidelines. Therefore, the treatment effect can be estimated only after accounting for confounding factors. 
Propensity score (PS) methods are a family of methods that is widely used for this purpose. Although they are all based on the estimation of the a posteriori probability of treatment assignment given patient covariates, they estimate the treatment effect from different statistical points of view and are, thus, relatively hard to compare.
In this work, we propose a simulation experiment in which a hypothetical cohort of subjects is simulated in seven scenarios of increasing complexity of the associations between covariates and treatment, but where the two main definitions of treatment effect (average treatment effect, ATE, and average effect of the treatment on the treated, ATT) coincide. Our purpose is to compare the performance of a wide array of PS-based methods (matching, stratification, and inverse probability weighting) in estimating the treatment effect and their robustness in different scenarios. 
We find that inverse probability weighting provides estimates of the treatment effect that are closer to the expected value by weighting all subjects of the starting population. Conversely, matching and stratification ensure that the subpopulation that generated the final estimate is made up of real instances drawn from the starting population, and, thus, provide a higher degree of control on the validity domain of the estimates. 
}

\begin{document}
\thispagestyle{myheadings}
\pagestyle{myheadings}
\markright{\tt Proceedings of CIBB 2024}

\section{\bf Introduction}
\label{sec:SCIENTIFIC-BACKGROUND}
In randomized controlled trials (RCTs), which are considered the gold standard to estimate the effect of treatments, the randomness of treatment assignment ensures that treatment status is not influenced by externals factor. This is not true for observational studies, where treatment allocation is usually influenced by subject characteristics, reimbursement policies, or existing guidelines, resulting in treated and untreated subpopulations that might be substantially different between each other.
Therefore, confounding factors must be taken into account before estimating treatment effects. 
This is usually done via propensity score (PS) methods, i.e., methods that aim to induce a quasi-equal distribution of covariates between treated and untreated subjects, trying to emulate the randomness of an RCT. Specifically, the PS is defined as the probability of treatment assignment conditional on a set of observed covariates and can be used as a metric for comparison between subjects. PS-based methods assume that individuals with similar PS values have, on average, similar covariates' distributions and can therefore be compared to estimate treatment effects. The most common PS-based methods are PS matching (PSM), PS stratification (PSS), and inverse probability weighting (IPW)~\cite{ps_methods1, ps_introduction}. These flavors of PS methods differ in their approach: PSM creates pairs or groups of treated and untreated subjects, which are then pooled together; PSS splits subjects into a fixed number of strata on the basis of the PS; and IPW creates a new virtual population by weighting each individual by a function of the PS.

The common target when analysing clinical data is the effect of a treatment, which is usually quantified by two estimands: the average treatment effect (ATE) and the average treatment effect on the treated (ATT). In RCTs, these two quantities are equal, since the distribution of covariates between treated and untreated is balanced, a condition that is not satisfied in observational studies due to confounding factors. PS-based methods do not target the same treatment effect. In particular, PSM attempts to estimate the ATT, because the matched pairs are taken based on the PS, which depends on the distribution of the treated, while PSS and IPW allow the estimation of both ATE and ATT~\cite{paper_extra}. As a result, in real-world scenarios, the direct comparison of PS-based methods is challenging.

To address this issue, we conducted an experiment with artificial data in which the ATE and ATT were equal by construction. Several scenarios were considered, characterized by different degrees of non-linearity and/or non-additivity in the relationship between the covariates and the treatment. The different PS-based methods were applied to each scenario and compared based on their ability to predict the ATE/ATT set at simulation time.

\section{\bf Data and Methods}
\label{sec:DATA-AND-METHODS}
\subsection{\bf Data generation}
Following what was described in \cite{ps_scenario}, we generated artificial data representing a hypothetical cohort study of 20,000 subjects characterized by a binary treatment $A$, a binary outcome $Y$, and ten covariates $W_i$, $i=1,\dots,10$. Covariates were generated from a multinomial normal distribution with zero mean, unit variance, and different correlation coefficients between its components. Covariates $W_1, W_3, W_5, W_6, W_8, W_9$ were dichotomized with a threshold of 0, while covariates $W_2, W_4, W_7, W_{10}$ were left continuous. 

Seven different scenarios (\textit{Scenario A-G}) were simulated by varying the non-linearity and/or non-additivity in the association between the covariates and the treatment (~\autoref{tab:scenarios}). The true PS models, which, in each scenario, predicted the probability of receiving the treatment $A$ given the set of covariates $W_i$, were defined by increasingly complex logistic regression models.
The simplest one (\textit{Scenario A}) assumed only linear associations and additive effects (~\autoref{eq:ps_A}), whilst the most complex one (\textit{Scenario G}) assumed three quadratic terms and ten two-way interaction terms involving confounders (~\autoref{eq:ps_G}).
\begin{table}[httb!]
\centering
\resizebox{\textwidth}{!} {
\begin{tabular}{lccc}
\toprule
 &
  \begin{tabular}[c]{@{}c@{}}Additive\\ (main effects only)\end{tabular} &
  \begin{tabular}[c]{@{}c@{}}Mild non-additivity\\ (three two-way interaction terms)\end{tabular} &
  \begin{tabular}[c]{@{}c@{}}Moderate non-additivity\\ (10 two-way interaction terms)\end{tabular} \\ 
  \midrule
\rowcolor{LightBlue}Linear (main effects only)                     & A & D & F \\
Mild non-linearity (one quadratic term)        & B & E & - \\
\rowcolor{LightBlue}Moderate non-linearity (three quadratic terms) & C & - & G \\
\bottomrule
\end{tabular}}
\caption{\textbf{Description of the scenarios A-G for data generation~\cite{ps_scenario}.}}
\label{tab:scenarios}
\end{table}

\begin{equation} \small
    \label{eq:ps_A}
    \begin{gathered}
    \textit{Scenario A:} \hspace{0.3cm} P[A=1|W_i]=(1+exp\{-(\beta_0 +\beta_1 W_1 +\beta_2 W_2 + \beta_3 W_3 +\beta_4 W_4 + \\ + \beta_5 W_5 + \beta_6 W_6 +\beta_7 W_7)\})^{-1}
    \end{gathered}
\end{equation}

\begin{equation} \small
    \label{eq:ps_G}
    \begin{gathered}
    \textit{Scenario G:} \hspace{0.3cm} P[A=1|W_i]=(1+exp\{-(\beta_0 +\beta_1 W_1 +\beta_2 W_2 + \beta_3 W_3 +\beta_4 W_4 + \\
    +\beta_5 W_5 +\beta_6 W_6 +\beta_7 W_7 +\beta_2W_2W_2+\beta_4 W_4 W_4 + \beta_7W_7W_7+\\
    +\beta_1\times 0.5\times W_1W_3+\beta_2\times0.7\times W_2W_4   + \beta_3\times 0.5 \times W_3 W_5 +\\
    + \beta_4\times 0.7\times W_4 W_6+\beta_5 \times 0.5 \times W_5W_7 
    + \beta_1\times 0.5 \times W_1W_6+ \\+ \beta_2\times 0.7 \times W_2W_3 + \beta_3\times 0.5 \times W_3W_4 +    \beta_4\times 0.5 \times W_4W_5 + \beta_5 \times 0.5\times W_5W_6)\})^{-1}
    \end{gathered}
\end{equation}

The outcomes $Y$ were also generated with a logistic regression model. Specifically, for all scenarios, the model used to simulate the outcome was:
\begin{equation} \small
\label{eq:outcome}
    \begin{gathered}
        P[Y=1|A, W_i] = (1+exp\{-(\alpha_0+\alpha_1 W_1+\alpha_2 W_2+\alpha_3 W_3
        + \alpha_4 W_4 +\alpha_5 W_8 +\\
        +\alpha_6 W_9+\alpha_7 W_{10}+\hat{\gamma_1} A)\})^{-1}
    \end{gathered}
\end{equation}

In summary, in all scenarios, covariates $W_1, W_2, W_3, W_4$ were associated with both the treatment and the outcome, covariates $W_5, W_6, W_7$ only with the treatment, and $W_8, W_9, W_{10}$ only with the outcome.

\subsection{\bf Methods}
The main PS-based methods were investigated: PSM, IPW, and PSS. 

PSM aims at creating two cohorts of individuals, respectively treated and untreated, having similar PS distributions. It explores the estimated PS values trying to match each treated subject with one or more untreated subjects on the basis of the PS similarity. This similarity can be limited by a caliper, which defines the maximum distance for which the matching is allowed. In this work, PSM was performed 1:1 without replacement with the nearest neighbor algorithm within a specified caliper equal to 10\% the pooled standard deviation of the estimated PS. To evaluate the stability of the estimates, a resampling scheme was implemented. It drew samples from the original dataset and for each sample, the PS model was estimated, matching was applied and the treatment effect was estimated.

IPW uses the PS to compute weights for all subjects which are defined as the inverse of the probability of receiving the treatment that the subject actually received: $w_i=1/{P(A_i|W^i_{1:10})}$, i.e., the inverse of the probability score for the treated, and of its ones' complement for the untreated. These weights are, then, used in a weighted estimator of the treatment effect. This procedure can be thought of as estimating the treatment effect in a virtual population in which the distribution of baseline covariates is independent of treatment assignment~\cite{ipw_paper}. Two extensions of IPW aiming at avoiding extreme values due to small PS values also exist: truncation (which works by saturating extreme weights laying in the lowest and highest 
percentiles) and stabilization (where the weights are set to: $w_i={P(A_i)}/{P(A_i|W^i_{1:10})}$). 

Finally, PSS splits all subjects into strata based on the estimated PS. The number of strata is usually set to five \cite{pss_strata}. The strata division can be performed in two ways. Subjects can be divided into equal-size strata such that each stratum has the same numerousness (PSS by quantile) or the range of estimated PS values can be divided into strata with equal width (PSS by PS value). In both techniques, the stratum-specific treatment effect is estimated within each stratum, then the average treatment effect is calculated as weighted average of the stratum-specific ones.

Our purpose was the comparison of these methods in estimating the ATE/ATT in the proposed scenarios. Let $Y^1$ and $Y^0$ be the outcome variable in the treatment and control groups, respectively, then the ATE is defined as:
\begin{equation} \small
    \label{eq:ATE}
    \text{ATE} = \text{E}[Y^1]-\text{E}[Y^0]
\end{equation}
and the ATT as:
\begin{equation} \small
    \label{eq:ATT}
    \text{ATT} = \text{E}[Y^1|A=1]-\text{E}[Y^0|A=1]
\end{equation}

In our simulation, the ATE and ATT were equal and represented by the coefficient $\gamma_1$ in ~\autoref{eq:outcome}, which was constant and equal to $-0.4$. 
Following standard practices in the field of observational studies, once a PS-based method was applied, the ATE/ATT was estimated using a logistic regression model in which the only input variable was the treatment $A$:
\begin{equation} \small
    \label{eq:gamma}
    \begin{gathered}
        P[Y=1|A] = (1+exp\{-(\hat{\gamma_0} + \hat{\gamma_1} A)\})^{-1}
    \end{gathered}
\end{equation}
We simulated a hypothetical cohort of 20,000 individuals and applied PSM, IPW, and PSS, and their variations. We repeated the application of the PS-based methods 1000 times by iteratively resampling 70\% of the starting cohort. A well-performing PS-based method would, then, estimate a value of $\gamma_1$ as close as possible to the true value of -0.4.

\section{\bf Results and Discussion}
\label{sec:RESULTS}
\begin{figure}[t]
\centering
\includegraphics[width=0.5\textwidth]{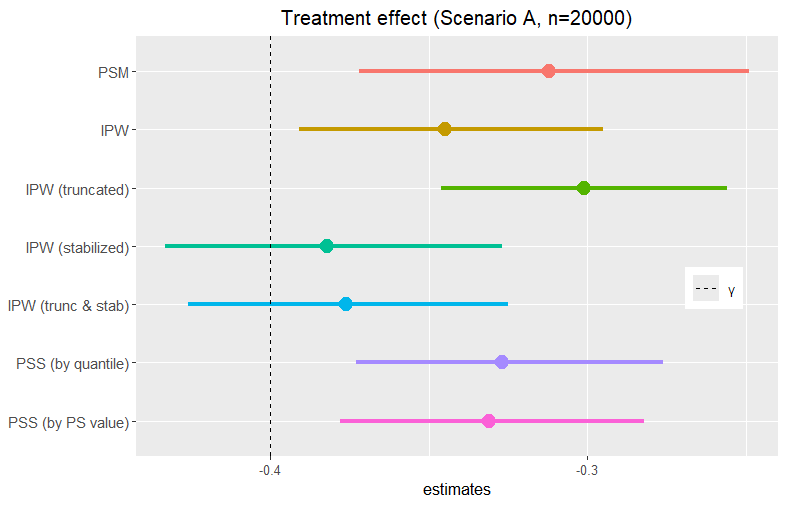}
\centering \includegraphics[width=0.5\textwidth]{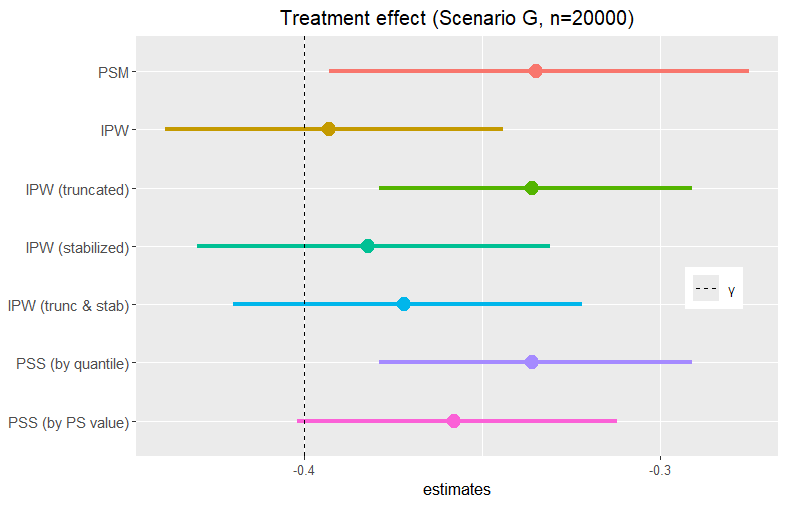}
\caption{{\bf Estimated treatment effect for scenarios A and G}. For each investigated method, the mean value and its 95\% confidence interval is plotted. Left and right panel shows the results respectively of scenario {\it A} and {\it G}.}
    \label{fig:results}
\end{figure}

We reported the average treatment effect estimate and its 95\% confidence interval for each method and scenario in \autoref{tab:RESULTS}. For illustrative purposes, we also plotted the results for scenarios {\it A} and {\it G} in \autoref{fig:results}.
We found that weighting methods provided estimates closer to the true value compared to matching and stratification methods. In particular, IPW with stabilized weights was the most accurate and stable method even in the most complex scenarios, where it estimated a treatment effect very close to the true one (e.g., exactly -0.4 in scenarios {\it C} and {\it D}). Instead, IPW with truncated weights had lower performance, while the combination of truncation and stabilization did not improve the estimates compared to stabilization alone. We hypothesize that this may have happened because, since truncation saturates the most extreme values, the distribution of PS values is modified and some individuals are assigned a weight that can be quite far from the intended one. This modification, then, introduces a bias in the model for estimating the treatment effect, which results in a higher estimation error.
\begin{table}[httb!] 
\centering
\resizebox{\textwidth}{!} {
\begin{tabular}{cccccccc}
\toprule
 & \multicolumn{7}{c}{\textbf{Scenario}} \\
\textbf{Method} & \textbf{A} & \textbf{B} & \textbf{C} & \textbf{D} & \textbf{E} & \textbf{F} & \textbf{G} \\
\midrule
\rowcolor{LightBlue} PSM & \begin{tabular}[c]{@{}c@{}}-0.312\\ {[}-0.372, -0.249{]}\end{tabular} & \begin{tabular}[c]{@{}c@{}}-0.297\\ {[}-0.358, -0.240{]}\end{tabular} & \begin{tabular}[c]{@{}c@{}}-0.366\\ {[}-0.420, -0.310{]}\end{tabular} & \begin{tabular}[c]{@{}c@{}}-0.346\\ {[}-0.413, -0.283{]}\end{tabular} & \begin{tabular}[c]{@{}c@{}}-0.374\\ {[}-0.433, -0.316{]}\end{tabular} & \begin{tabular}[c]{@{}c@{}}-0.332\\ {[}-0.397, -0.268{]}\end{tabular} & \begin{tabular}[c]{@{}c@{}}-0.335\\ {[}-0.393, -0.275{]}\end{tabular} \\
IPW & \begin{tabular}[c]{@{}c@{}}-0.345\\ {[}-0.391, -0.295{]}\end{tabular} & \begin{tabular}[c]{@{}c@{}}-0.317\\ {[}-0.366, -0.271{]}\end{tabular} & \begin{tabular}[c]{@{}c@{}}-0.379\\ {[}-0.423, -0.339{]}\end{tabular} & \begin{tabular}[c]{@{}c@{}}-0.373\\ {[}-0.424, -0.323{]}\end{tabular} & \begin{tabular}[c]{@{}c@{}}-0.371\\ {[}-0.422, -0.324{]}\end{tabular} & \begin{tabular}[c]{@{}c@{}}-0.376\\ {[}-0.424, -0.327{]}\end{tabular} & \begin{tabular}[c]{@{}c@{}}-0.393\\ {[}-0.439, -0.344{]}\end{tabular} \\
\rowcolor{LightBlue} \begin{tabular}[c]{@{}c@{}}IPW\\ (truncated)\end{tabular} & \begin{tabular}[c]{@{}c@{}}-0.301\\ {[}-0.346, -0.256{]}\end{tabular} & \begin{tabular}[c]{@{}c@{}}-0.280\\ {[}-0.323, -0.237{]}\end{tabular} & \begin{tabular}[c]{@{}c@{}}-0.349\\ {[}-0.391, -0.309{]}\end{tabular} & \begin{tabular}[c]{@{}c@{}}-0.321\\ {[}-0.367, -0.275{]}\end{tabular} & \begin{tabular}[c]{@{}c@{}}-0.331\\ {[}-0.374, -0.286{]}\end{tabular} & \begin{tabular}[c]{@{}c@{}}-0.327\\ {[}-0.368, -0.281{]}\end{tabular} & \begin{tabular}[c]{@{}c@{}}-0.336\\ {[}-0.379, -0.291{]}\end{tabular} \\
\begin{tabular}[c]{@{}c@{}}IPW \\ (stabilized)\end{tabular} & \begin{tabular}[c]{@{}c@{}}-0.382\\ {[}-0.433, -0.327{]}\end{tabular} & \begin{tabular}[c]{@{}c@{}}-0.347\\ {[}-0.402, -0.298{]}\end{tabular} & \begin{tabular}[c]{@{}c@{}}-0.401\\ {[}-0.449, -0.358{]}\end{tabular} & \begin{tabular}[c]{@{}c@{}}-0.400\\ {[}-0.456, -0.346{]}\end{tabular} & \begin{tabular}[c]{@{}c@{}}-0.396\\ {[}-0.451, -0.344{]}\end{tabular} & \begin{tabular}[c]{@{}c@{}}-0.396\\ {[}-0.448, -0.342{]}\end{tabular} & \begin{tabular}[c]{@{}c@{}}-0.382\\ {[}-0.430, -0.331{]}\end{tabular} \\
\rowcolor{LightBlue} \begin{tabular}[c]{@{}c@{}}IPW \\ (trunc \& stab)\end{tabular} & \begin{tabular}[c]{@{}c@{}}-0.376\\ {[}-0.426, -0.325{]}\end{tabular} & \begin{tabular}[c]{@{}c@{}}-0.339\\ {[}-0.387, -0.291{]}\end{tabular} & \begin{tabular}[c]{@{}c@{}}-0.395\\ {[}-0.443, -0.350{]}\end{tabular} & \begin{tabular}[c]{@{}c@{}}-0.399\\ {[}-0.451, -0.348{]}\end{tabular} & \begin{tabular}[c]{@{}c@{}}-0.393\\ {[}-0.444, -0.342{]}\end{tabular} & \begin{tabular}[c]{@{}c@{}}-0.395\\ {[}-0.442, -0.341{]}\end{tabular} & \begin{tabular}[c]{@{}c@{}}-0.372\\ {[}-0.420, -0.322{]}\end{tabular} \\
\begin{tabular}[c]{@{}c@{}}PSS\\ (by quantile)\end{tabular} & \begin{tabular}[c]{@{}c@{}}-0.327\\ {[}-0.373, -0.276{]}\end{tabular} & \begin{tabular}[c]{@{}c@{}}-0.288\\ {[}-0.334, -0.245{]}\end{tabular} & \begin{tabular}[c]{@{}c@{}}-0.358\\ {[}-0.402, -0.317{]}\end{tabular} & \begin{tabular}[c]{@{}c@{}}-0.343\\ {[}-0.393, -0.294{]}\end{tabular} & \begin{tabular}[c]{@{}c@{}}-0.332\\ {[}-0.38, -0.286{]}\end{tabular} & \begin{tabular}[c]{@{}c@{}}-0.351\\ {[}-0.396, -0.303{]}\end{tabular} & \begin{tabular}[c]{@{}c@{}}-0.336\\ {[}-0.379, -0.291{]}\end{tabular} \\
\rowcolor{LightBlue} \begin{tabular}[c]{@{}c@{}}PSS\\ (by PS value)\end{tabular} & \begin{tabular}[c]{@{}c@{}}-0.331\\ {[}-0.378, -0.282{]}\end{tabular} & \begin{tabular}[c]{@{}c@{}}-0.297\\ {[}-0.341, -0.252{]}\end{tabular} & \begin{tabular}[c]{@{}c@{}}-0.366\\ {[}-0.412, -0.324{]}\end{tabular} & \begin{tabular}[c]{@{}c@{}}-0.350\\ {[}-0.400, -0.300{]}\end{tabular} & \begin{tabular}[c]{@{}c@{}}-0.351\\ {[}-0.399, -0.302{]}\end{tabular} & \begin{tabular}[c]{@{}c@{}}-0.349\\ {[}-0.392, -0.301{]}\end{tabular} & \begin{tabular}[c]{@{}c@{}}-0.358\\ {[}-0.402, -0.312{]}\end{tabular}
\\ 
\bottomrule
\end{tabular}
}
\caption{\textbf{Treatment effect estimated on the 1000 resamplings of the 20,000-subject cohort}. For each method and scenario, the mean and 95\% confidence interval are reported. \label{tab:RESULTS}}
\end{table}

We also found that PSM produces biased, but very stable, estimates of the treatment effect averaging around -0.3 across scenarios. This was likely due to the fact that the PSM does not estimate the ATT in the whole population but only in the subgroup obtained in the matching and after the application of the caliper, which, in turn, results in a distortion of the distribution of the treated relative to the one that would have allowed us to estimate the expected value of -0.4. However, this consideration does not invalidate the PSM results because, unlike IPW, which amounts to creating a new, virtual population, the PSM population on which the treatment effect is calculated is fully known and \textit{real} as it is simply a subset of the original one. In other words, the domain of validity of PSM, i.e., in our case, the matched subpopulation in which the ATT is -0.3 instead of -0.4, can always be verified and is ensured to comprise only real instances drawn from the observed data and no virtual patients. Conversely, despite how well IPW estimated the ATE (which, here, was equal to the ATT) in our experiments, such guarantees cannot be given due to the effect of weighting.

Finally, PSS exhibited intermediate performance compared to IPW and PSM, with the goodness of the estimates getting closer to IPW's in more complex scenarios. Considering the two ways of dividing subjects into strata, splitting by PS value yielded better estimates than splitting by quantile.

A priori, it would have been natural to think that the PS-based methods would have more difficulties in dealing with the non-linear and/or non-additive relationships of the more complex scenarios. However, remarkably, in our experiments, all PS-methods provided relatively stable estimates of the treatment effect even in the most complex scenario, as shown in~\autoref{fig:results}. Overall, IPW stabilized had the best performances across all scenarios, while PSM and PSS varied slightly more. While in the simpler scenarios, PSS provided better estimates than PSM, in the more complex ones the two methods became comparable in their performance. 

\section{\bf Conclusion}
\label{sec:CONCLUSIONS}
In this work, we proposed an experiment in which PS-based methods, such as PSM, PSS and IPW, where applied to a simulated dataset representing a hypothetical cohort of 20,000 individuals. The experiment was designed such that the investigated methods could be directly compared, because the target ATE and ATT were equal. Our purpose was to evaluate their performance in estimating the treatment effect in seven scenarios differing in the degrees of non-linearity and/or non-additivity association between the covariates and the treatment. We found that IPW implemented with stabilized weights had the best performance in estimating the treatment effect compared to PSS and PSM in all scenarios.  

In a sense, however, this superiority in terms of quality of the estimates comes at a cost. Indeed, IPW  does not directly model the original population, but a new, weighted version thereof. This results in a lower degree of control on the domain of validity of the final estimate compared to PSM, where the subjects involved in the final estimate are all guaranteed to be unaltered members of the starting cohort.
The intermediate performance of PSS also appears to be consistent with the aforementioned trade-off between estimate accuracy and control over the population. In fact, PSS has the advantages of considering all the unweighted individuals of the population, without discarding any like PSM would, and to not modify their weights in the treatment effect estimation like IPW. However, it does not eliminate confounding factors as strongly as either method.

On the basis of our results, we suggest the use of weighting methods, especially IPW implemented with stabilized weights, when the purpose of the analysis is to obtain a good estimate of the ATE in situation where an accurate specification of the domain of validity is of secondary importance. On the contrary, if tighter control is needed (e.g., because the individual-level values of some covariates are relevant to the decision-making process informed by the experiment), PSM or PSS implemented with the strata division based on the PS values might be preferable. \\
While simulations study exist in the literature, compared to existing works in this field~\cite{ps_methods2, paper_extra2}, ours tests a wider variety of techniques, both in the weighting and matching categories.
Future developments of this study will include the validation of our results on different simulation settings, especially investigating regression and survival models of the outcome. Furthermore, we will explicitly test different mathematical formulations of the treatment effect (e.g., linear effect, risk ratio, risk difference).
Future work will also include the evaluation of other approaches for treatment effect estimation, such as G-computation and Targeted Maximum Likelihood Estimation. 

\section*{\bf Conflict of interests}
\label{sec:CONFLICT-OF-INTERESTS}
All authors declare that they have no conflicts of interest.

\section*{\bf Funding}
\label{sec:FUNDING}
This work was funded by the REDDIE (Real-world evidence for decisions in diabetes) project. The REDDIE project has received funding from the European Union's Horizon 2022 research and innovation programme under grant agreement No. 101095556. Views and opinions expressed are however those of the author(s) only and do not necessarily reflect those of the European Union or European Health and Digital Executive Agency (HADEA). Neither the European Union nor the granting authority can be held responsible for them. This work has received funding from the UK research and Innovation under contract number 101095556.

\footnotesize
\bibliographystyle{unsrt}
\bibliography{main.bib} 
\normalsize

\end{document}